\pdfoutput=1 
\documentclass{JINST}
\usepackage[normalem]{ulem}
\usepackage{amsmath}

\title{A versatile and light-weight slow control system for small-scale applications}

\author{Philipp Zappa, Lukas B\"utikofer, Daniel Coderre, Basho Kaminsky, Marc Schumann, Moritz von Sivers \\
\llap{}Albert Einstein Center for Fundamental Physics, Universit\"{a}t Bern, 3012 Bern, Switzerland\\
  E-mail:  \email{philipp.zappa@bluewin.ch}, \email{marc.schumann@lhep.unibe.ch}}

\abstract{We present an open source slow control system for small and medium scale projects. Thanks to its modular and flexible design, where the various instruments are read and controlled by independent plugins, Doberman (Detector OBsERving and Monitoring ApplicatioN) can be quickly adapted for many applications, also making use of existing code or proprietary components. The system uses a  SQL~database to store the data from the instruments and provides an online application to display and browse through the data. It allows the modification of device settings while the program is running and features a protocol to handle exceptions, including the automated distribution of alarm messages. We present two case studies from astroparticle physics, on which Doberman is successfully deployed: a low-background screening facility installed in a remote underground laboratory and a detector R\&D platform using cryogenic liquid xenon.}

\keywords{Detector control systems (detector and experiment monitoring and slow-control systems, architecture, hardware, algorithms, databases), Control and monitor systems online}

\begin{document}

\section{Introduction}
\label{sec::introduction}

Slow control systems are used to monitor a wide variety of experimental apparatus. This is especially critical for experiments that must operate autonomously over long periods of time or are installed in remote locations.
In contrast to the ``fast'' data acquisition system (DAQ), which is designed to read out the scientific data from the detector once they occur at usually rather high rates, the slow control system usually reads and processes auxiliary sensors and devices at regular intervals, ranging from seconds to hours. Typical slow control parameters are high voltages, temperatures, pressures, gas flows, but also include digital on/off states, e.g., of crates.

Typical slow control systems in particle physics, as for example described in~\cite{ref:atlas,ref:xenon,ref:reno}, set the parameters of various instruments (``configuration'' and ``control''), record and store the measurements of the instruments' sensors for online and offline use (``monitor''), and provide an automated feedback  to the user in case of parameters falling outside a pre-defined range (``exception''). The latter is of particular importance for safety-relevant applications.

While some large projects employ industrial SCADA (supervisory control and data acquisition) standards and systems~\cite{ref:atlas}, 
similar efforts can usually not be afforded for considerably smaller laboratory projects with their limited scientific scope, operation time and manpower. A full SCADA performance, where complex modification of a detector's state are managed by the system, is often also not required. Open-source slow control solutions, such as EPICS (Argonne)~\cite{ref::epics}, NOMAD (ILL)~\cite{ref:nomad} or Midas (PSI/TRIUMF)~\cite{ref:midas} exist for large~\cite{ref::d0,ref::babar} to medium-sized experiments (e.g.,~\cite{ref:meg}) or research centers with many experiments of similar instrumentation. However, also these are often too complex for small R\&D projects with only a few different instruments. As a consequence, such systems are often either operated without slow control, or with custom solutions with limited functionality.

To fill this gap we have developed Doberman (Detector OBsERving and Monitoring ApplicatioN). It is a simple and extendable sopen source slow control system designed for small projects~\cite{ref::download}. Its lightweight core installation provides basic functionality, such as communications protocols, data storage, and data visualization, while the instrument readout is realized via independent plugins. This extendable plugin structure is well-suited for small projects which are typically installed, operated and eventually upgraded in stages. The plugins are independent of the Doberman core, which allows the easy integration of pre-existing code for instruments into the common framework. Doberman was designed to ensure that malfunctioning of individual devices or plugins does not affect the stability and integrity of the system.

The article is structured as follows: the Doberman slow control system, including its alarm functionality and online data display unit, is described in Section~\ref{sec:Software}. We demonstrate the successful application of the system to two different detector systems, a low-background germanium spectrometer installed in a remote underground laboratory and a liquid-xenon detector setup for detector R\&D, in Section~\ref{sec:Examples} and conclude in Section~\ref{sec:Conclusion}.

\section{The Doberman slow control system}  \label{sec:Software}

\begin{figure}[b]
\centering
\includegraphics[width=0.99\columnwidth]{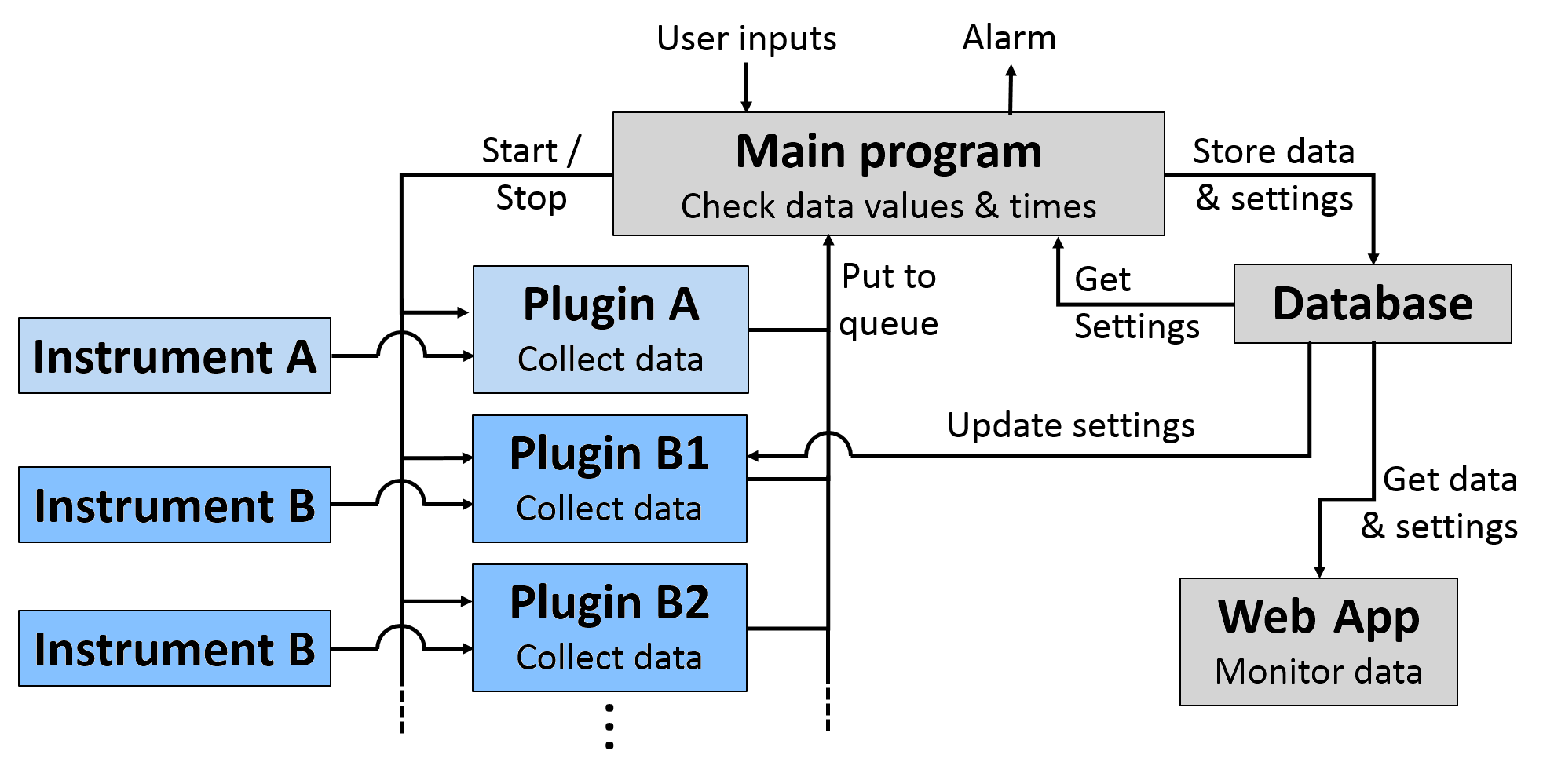}
\caption{Doberman software scheme and data flow. The number of plugins can be extended considerably, matching the requirements of the project. The user operates the software through the ``Main program'' and can observe the collected data in real time via a web application (``Web App''). User inputs include the start and stop commands, as well as all settings for Doberman and the plugins. Optionally, the plugins can communicate with the database to update device-specific settings while the system is running. Identical plugins can be used for identical instruments (``Instrument B''), but have to be named differently. 
\label{fig:Dataflow} }
\end{figure}

An overview of the Doberman system, with its sub-components projects and the information flow is shown in Figure~\ref{fig:Dataflow}.  It consists of a main program, which handles the data and organizes the communication, and a plugin for each instrument, which provides the slow control data. An instrument (e.g., a temperature controller) can deliver several individual parameters to the system (e.g., set temperatures, measured temperatures, proportional-integral-derivative (PID) control parameters, etc.).
A data queue is used for the communication between the main program and the plugins, which depend on the use case and have to be written by the user. For fast and flexible access to the data, it is stored in a PostgreSQL database after its validity has been verified by the main program, which also raises alarms in case of exceptions. A customizable web-based monitor allows the online display of the data. Doberman was developed in Python and is currently operated on Linux (Ubuntu), but in general it is cross-platform compatible.

The Doberman main program is isolated from the plugins and resilient to plugin crashes. Communication with the plugins (see Section~\ref{sec:Plugin_structure}) to set parameters and processing/storage of the delivered data including error handling is performed in a separated thread, which can be restarted by the main process in case of instabilities. In such case, an alarm is raised. The plugins can be started/stopped and their settings can be changed without stopping the program. The settings are either defined via shell prompts or taken directly from the database (default).

The data from the first-in-first-out (FIFO) queue is stored in a database with a time stamp and a status parameter, which describes whether an error occurred during parameter acquisition. In addition, each datapoint is compared to an optionally defined parameter range. The users are automatically notified (by email or SMS) if an exception is met, e.g., if a parameter is out of range, its status parameter faulty, or if a device does not deliver data at the expected transmission rate. Whenever it is idle, the main program checks the database for updates of the user-defined Doberman and alarm settings.

\subsection{Instrument plugins} \label{sec:Plugin_structure}

\begin{figure}[b!]
\begin{minipage}[]{0.4\textwidth}
\begin{center}
\includegraphics[width=0.99\textwidth]{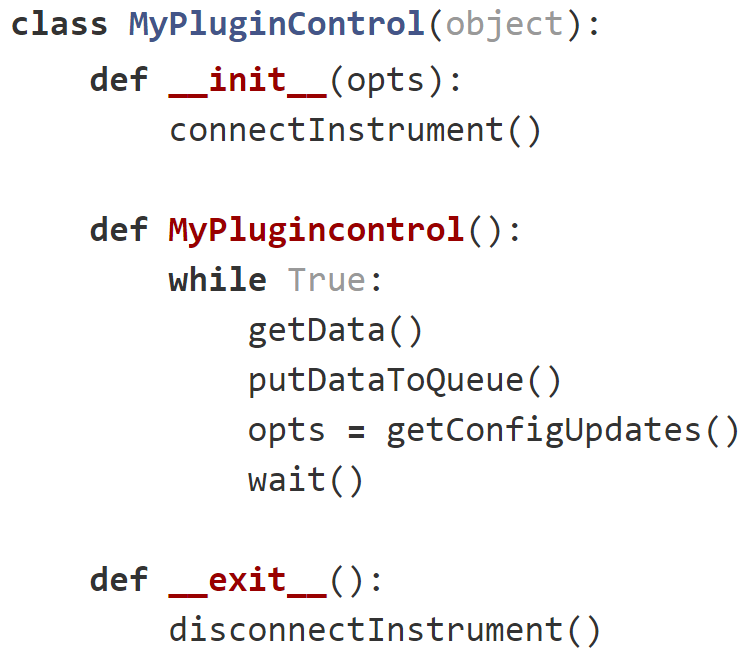}
\end{center}
\end{minipage}
\hfill 
\begin{minipage}[]{0.58\textwidth}
\vspace{-0.5cm}
\caption{Pseudo-code illustrating the minimal plugin requirements. When Doberman imports the constructor, the class {\tt MyPluginControl} is loaded and {\tt \_\_init\_\_(opts)} called. The options {\tt opts} provide the plugin settings (e.g., connection to instrument, setpoints, etc.). {\tt MyPluginControl()} and {\tt \_\_exit\_\_} are called to start (in a thread) and close the plugin, respectively. All other functions shown here ({\tt connectInstrument()}, {\tt getData()}, etc.) are generic examples and call other code or programs. The optional function {\tt getConfigUpdates()} updates the plugin settings during operation. \label{fig:Plugin_example}}
\end{minipage}
\end{figure}

Instruments interact with the main program through plugins. For maximal flexibility, Doberman imposes only minimal requirements for the plugins, namely some naming conventions and a limited set of generic functions for instrument initialization, start, stop and data transfer, as illustrated in Figure~\ref{fig:Plugin_example}. Any other functionality is to be defined by the user. The plugin must be written in Python, however, a slim Python wrapper can be used as interface to another program written in any programming language or to software provided by the instrument manufacturer. Each instrument requires a unique plugin that collects data independently from all other devices at a pre-defined rate and delivers it to the queue. Instrument control parameters, e.g., setpoints, can be updated through the plugin. This allows the user to change the parameters via the main program while the plugin fetches the new settings from the database. Plugins can also be configured to perform continuous, active instrument control, such as a PID feedback loop. 

All activated plugins are imported sequentially once Doberman is started. After their initialization with connection information (protocol, rate) and setpoints, the data transfer queue is established. From this point, the plugin runs with the initial settings, completely independently from the Doberman main program until it is eventually terminated. To update the configuration while the slow control system is running, the plugin has to be equipped with an optional direct interface to the database to periodically check for modified settings.

The stability of the main program with the exception handling is supported by several measures: (i) data transfer via one common queue, (ii) a minimal number of commands between the plugins and the main program, (iii) outsourcing of the time-consuming and error-prone processes such as hardware control and data collection to the plugins. The plugin structure also allows the addition of new features without any modification of the Doberman core, which facilitates code maintenance.

Additional safety features can be easily established via plugins as well. Two realized examples are plugins to monitor the available memory space and to check the integrity of the network connection. The latter could result in the potentially dangerous situation where an exception state cannot be reported to the users. The safety plugin constantly checks the network connection between two or more access points, which are preferentially located at different physical routers.

\subsection{Data storage and display} \label{sec:data_storage}

\begin{figure}[h!]
\centering
\includegraphics[width=0.99\columnwidth]{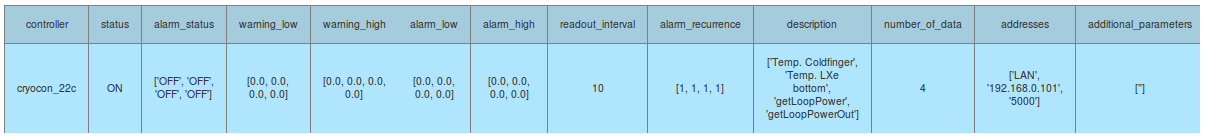} 
\caption{Example of the settings table for a cryogenic controller plugin, as stored in the Doberman database. The information can be easily accessed via an online monitor application. 
\label{fig:Webapp2} }
\end{figure}

The instrument's data it is stored in a PostgreSQL~\cite{ref:PostgreSQL} database. A database table is generated for each plugin, also containing timestamps and additional status parameters to store possible error conditions. As the maximum table size is 32\,TB~\cite{ref:PostgreSQL_Size}, the total size of the database is generally rather unlimited and depends mainly on the computing hardware. Table~\ref{tab:Database} shows such a data table for an example plugin. The settings for the various components (Doberman main program, exceptions/alarms, instrument plugins) and their history are also stored in database tables. As an example, the settings table of an existing instrument plugin is shown in Figure~\ref{fig:Webapp2}. 

\begin{table}[b]
\centering
\caption{Data table for the example plugin {\tt MyPlugin} which delivers the values of 3~different parameters in a single readout cycle every 30~seconds. In the database these values are stored as an array. The  ``Status''-array indicates an error condition ($0$:~good data, $-1$~no connection, $-2$:~no status, $1\ldots100$: instrument specific warnings/alarms, e.g., under range, sensor error, etc.). \label{tab:Database}}
\vspace{0.2cm}
\footnotesize
\begin{tabular}{|c|c|c|}
\hline
\multicolumn{3}{|c|}{\tt Data\_MyPlugin}  \\
\hline
\, \, \, {\bf Date} \,\, | \,\, {\bf Time} & {\bf Data} & {\bf Status} \\
\hline
'2016-04-11 | 11:36:37' & [0.12, 2.0, 14.123] & [0, 0, 0] \\
'2016-04-11 | 11:37:07' & [0.11, 2.0, 14.133] & [0, 0, 0] \\
'2016-04-11 | 11:37:37' & [0.11, 2.0, 14.139] & [0, 0, 0] \\ \hline
\end{tabular}
\end{table}

To avoid the risk of accidental data loss, the database entries cannot be deleted via Doberman. If necessary, it has to be done manually. Since the maximum table size is significantly larger than the total amount of data expected to be generated over the entire lifetime of typical Doberman applications (see examples in Section~\ref{sec:Examples}), this is currently not considered to be a relevant issue. In general, the system can be extended to allow for long-term data storage in a file system or for continuous data removal.

\begin{figure}[tbh]
\centering
\includegraphics[width=0.99\columnwidth]{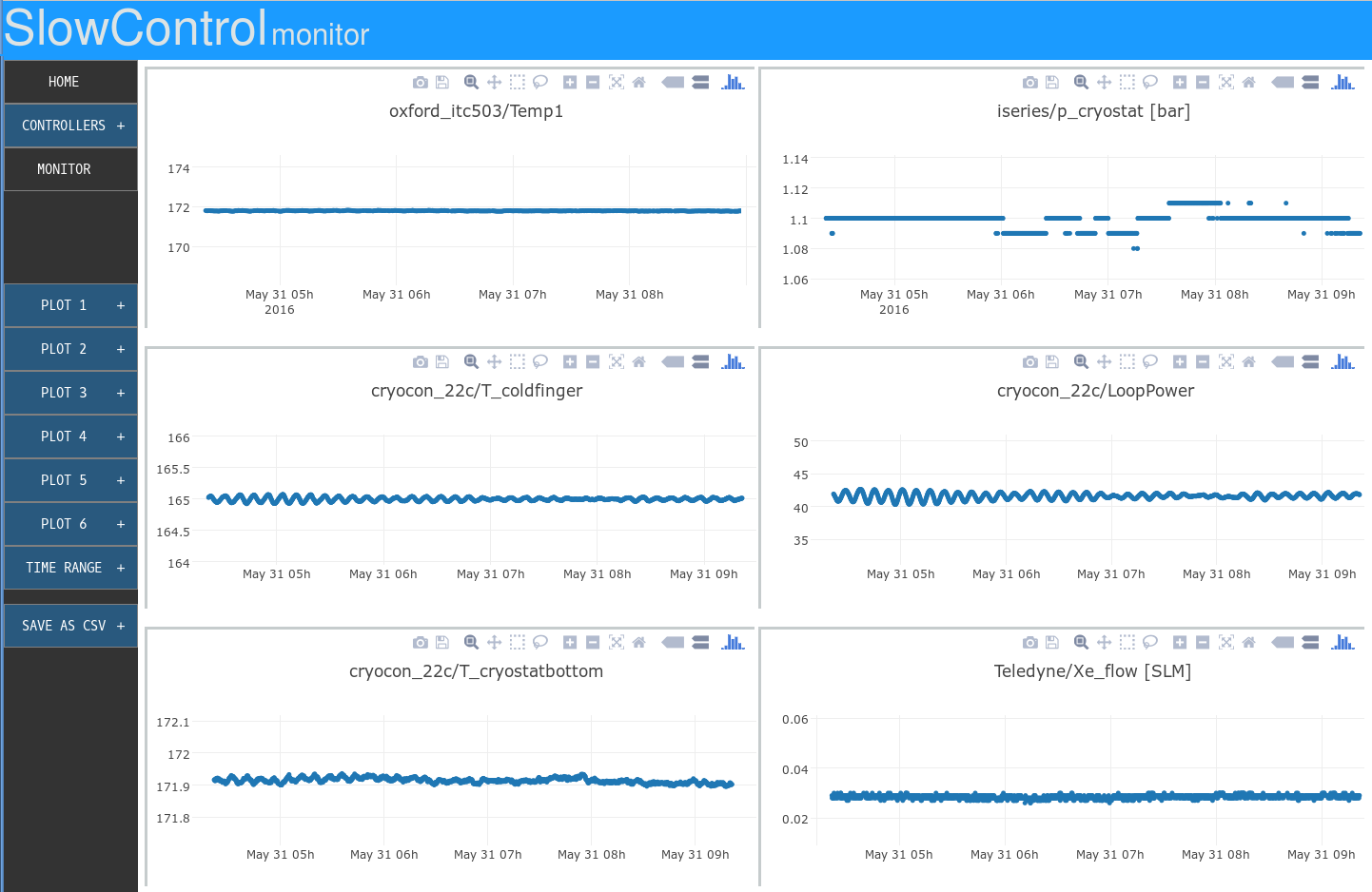}
\caption{The slow control display application runs in a standard web browser. The example (see also Section~3.2) shows six parameters (3~temperatures, 1~pressure, 1~power output and 1~gas flow) read from four different instruments. 17~parameters are monitored in total.  \label{fig:Webapp} }
\end{figure}

All data stored in the database can be displayed in real time by means of a web application, see Figures~\ref{fig:Webapp2} and~\ref{fig:Webapp}. The application's front page shows up to six panels with the time evolution of individual parameters, which can be selected and arranged by the user. The time interval as well as the vertical zoom-level can be adjusted directly on the panels. Export of the raw data in CSV~format as well as storing the plots as image files is also featured.

The display application can be accessed via the a standard web browser. It currently only allows the display and extraction of the data -- all inputs have to be performed on the host. However, if it becomes necessary for a specific application, it is straightforward to make the application online accessible and interactive, to deliver user inputs to the main program.

\subsection{Exception handling} \label{sec:alarm_handling}

In Doberman, an exception occurs when (i) a slow control parameter falls outside a pre-defined range, (ii) a parameter, device or plugin cannot be accessed or reports an error, (iii) the main program thread storing and checking the data failed or (iv) there is pile-up in the data queue. Exceptions are also raised by the monitoring plugins introduced in Section~\ref{sec:Plugin_structure} when (v) the database is full or (vi) the network is down. Similar to~\cite{ref:xenon}, Doberman distinguishes between two different types of exceptions, ``alarms'' and ``warnings'', which are to be defined by the user and lead to different notification mechanisms. 

In the standard configuration, warning conditions issue an email to a list of pre-defined users, which contains information on the exception plus additional information such as the parameter's time evolution and an history of alarm and warnings. The alarm condition additionally sends a SMS to a list of users. The distribution currently works over commercial email-to-SMS host platforms. However, in general also other ways are possible, such as the direct login to a SMS server or the use of a mobile phone SIM card within the slow control computer, which would maintain the alarm functionality also in case of network losses.  
The exception conditions, e.g., the alarm and warning thresholds, the number of consecutive data points exceeding the thresholds before a warning/alarm is triggered, the frequency of alarm messages if the problem remains, etc.,~as well as the user contact lists are set via the main program.

\subsection{Performance} \label{sec:Performance}

Doberman was designed to be used with small or medium-sized projects, with a limited number of instruments being read at frequencies ranging from seconds up to hours. Therefore, a rather small overall data rate is expected (see the examples in Section~\ref{sec:Examples} for typical numbers). We have tested on a standard computer (Intel Core i7-4770, quad-core 3.4\,GHz, 8\,GB RAM) that Doberman is able to process $>$2500~parameters delivered to the queue by 50~instruments (plugins) per second. The processing frequency remains unaffected even if every parameter raises a warning condition or if the format of every status parameter is incorrect. In general, the average input data rate needs to remain below this maximal processing frequency to ensure that updated slow control settings are regularly fetched from the database by the main program. A warning message can be generated if a maximal queue length is reached. We conclude that Doberman can safely handle an amount of parameters per second which is more than two orders of magnitude higher than required for typical small-scale applications (see examples in Section~\ref{sec:Examples}).

Doberman's processing capabilities could be improved by process threading and further optimization of the database usage. The slow control computer also plays a role for the overall performance, for example by limiting the number of ports to connect instruments (e.g., USB, RS-232, LAN) or by introducing stability problems once many plugins are running simultaneously. In future versions of Doberman, this could be improved by spreading the system over several machines, replacing the queue communication by a socket or database-centered communication.

\section{Example applications} \label{sec:Examples}

Doberman is currently successfully used for the slow control of two projects related to astroparticle physics and rare event searches. These are a low-background screening facility, described in Section~\ref{subsec:GeMSE}, and a small-scale R\&D platform which uses cryogenic liquid xenon as detector material, presented in Section~\ref{subsec:LabTPC}. To illustrate possible applications, we detail the various hardware components used in these projects and present measurements of the typical memory usage per day.

\subsection{Application 1: The GeMSE low-background screening facility} \label{subsec:GeMSE}

GeMSE (Germanium Material and meteorite Screening Experiment) is a low-radioactive background screening facility based on a high-purity germanium (HPGe) detector~\cite{ref:gemse,ref::gemse_instr}. It's science program is centered around the selection of construction materials and components for rare event search experiments, e.g., for the dark matter search projects XENON~\cite{ref:xenon1t} and DARWIN~\cite{ref:darwin}, and the identification of short-lived cosmogenic isotopes in meteorites~\cite{ref:meteorites}. GeMSE is installed at the Vue-des-Alpes underground laboratory~\cite{ref::vda}, located in the Swiss Jura mountains close to Neuch\^{a}tel at a depth of $\sim$600\,m water equivalent. This laboratory is about 1\,hour away from the University of Bern and is usually visited by personnel only every 2-3~weeks. The operation of the facility therefore relies entirely on a slow control system that can be accessed remotely. 

The HPGe detector needs to be operated at liquid nitrogen (LN$_2$) temperature, and the facility is equipped with an automated LN$_2$-refilling system. In order to avoid damage to the detector, it is important that its bias voltage is being turned off in case the cooling fails. The detector is installed inside a massive shield, which is purged with nitrogen (N$_2$) boil-off gas to reduce Rn-induced background, and is additionally equipped with a muon veto based on plastic scintillator panels. All instruments and parameters read by the Doberman slow control are listed in Table~\ref{tab:gemse}. The data from the HPGe detector is recorded by means of a commercial MCA unit, which is also remotely accessible but not included in the Doberman framework.

\begin{table}[h!]
\centering
\caption{Instruments of the GeMSE HPGe screening facility. The commercial data acquisition system reading the HPGe crystal is not integrated into the slow control (SC) framework.
\label{tab:gemse}}
\vspace{0.2cm}
\footnotesize
\begin{tabular}{| l | l | p{4.1cm} | c | p{3cm} |}
\hline
{\bf Instrument} & {\bf Model}  & {\bf Measured Parameters}    & {\bf Port} &    {\bf Comments}     \\ \hline
HV supply & Iseg NHQ105m~\cite{ref:iseq} & 4: bias voltage for HPGe crystal, set voltage, current, status & RS-232  & voltage set via SC\\
Multi-purpose unit & Labjack U12~\cite{ref:Labjack} & 5: muon veto rate, status LN$_2$ sensor and valve, HPGe leakage current, lab temperature & USB &  \\
Radon monitor & RAD7~\cite{ref:RAD7} & 14: Rn concentration, air temperature and humidity, 11~detector parameters & RS-232 & one new Rn value/hour \\
Flow controller & MKS G-Series~\cite{ref:MKS Flowmeter} & 2: N$_2$ gas purge flow, instrument temperature & RS-232 &  flow set via SC \\
\hline
\end{tabular}
\end{table}

\begin{figure}[t]
\centering
\includegraphics[width=0.7\columnwidth]{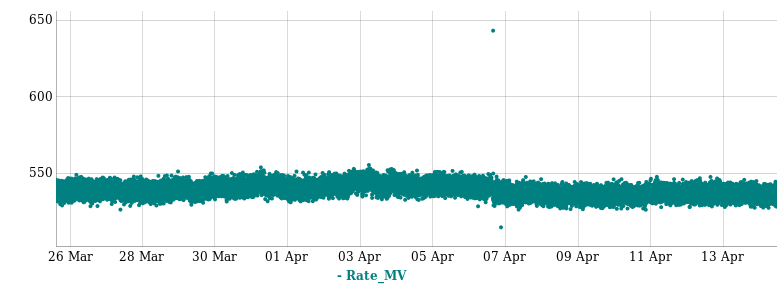}  
\caption{Muon veto detector rate (in Hz) of the GeMSE low-background screening facility, measured over a period of about 3\,weeks and displayed by the Doberman web application. \label{fig:gemse}}
\end{figure}

All devices listed in the Table are read out every 60~seconds by their respective plugins. Even at this rate, which could be reduced significantly for slow instruments such as the radon monitor, the amount of data in the database increases only by $\sim$1\,MB/day.
Doberman has been continuously monitoring GeMSE for 6~months, with only short interruptions for software updates and detector maintenance. As an example of a long-term slow control measurement, Figure~\ref{fig:gemse} shows the muon veto rate over a period of 3\,weeks, which is typical for the screening of a low-background material.

\subsection{Application 2: A liquid xenon test platform} \label{subsec:LabTPC}

Doberman is also used to monitor a detector R\&D platform which uses cryogenic liquid xenon as particle detection material. The astroparticle physics group at the University of Bern operates this system to develop and refine techniques for future dark matter detectors, e.g., DARWIN~\cite{ref:darwin}, and to perform tests for the XENON dark matter search project~\cite{ref:xenon1t}. The cryogenic liquid xenon is kept at $-$94$^\circ$C and $\sim$2\,bar absolute pressure in a vacuum-insulated double-wall cryostat. If these conditions can not be maintained, e.g., in case of a leak in the insulation vacuum or a sudden loss of the LN$_2$-based cooling power, the liquid xenon will evaporate and potentially lead to xenon losses and severe damage of the platform. Therefore, a slow control system is mandatory to continuously monitor various experimental parameters in order to avoid such situations.

\begin{table}[b!]
\centering
\caption{Instruments and plugins for the liquid xenon (LXe) R\&D platform at the University of Bern. Two identical process controllers to measure the xenon gas pressure $P_\textnormal{\tiny Xe}$ via pressure transducers and two universal transducer interfaces (UTI) monitoring the LXe level are read by two identical plugins, respectively. \label{tab:labsetup}}
\vspace{0.2cm}
\footnotesize
\begin{tabular}{| l | l | p{4.1cm} | c | c |}
\hline
{\bf Instrument} & {\bf Model}  & {\bf Measured Parameters}    & {\bf Port} &    {\bf Read period}     \\ \hline
Cryo controller & Cryo-Con 22c~\cite{ref:cryocon} & 4: LXe temperature (2),  heater output power (2) & LAN  & 10\,s \\
Cryo controller & Oxford ITC503~\cite{ref:oxford} & 4: Temperatures & RS-232 & 10\,s \\
Temp.~sensor & TEMPerNTC~\cite{ref:temperNTC} & 1: lab temperature & USB & 10\,s \\
Process controller & Newport i3200~\cite{ref:iseries} & 1: $P_\textnormal{\tiny Xe}$ (low) via transducer & RS-232 & 10\,s \\
Process controller & Newport i3200~\cite{ref:iseries} & 1: $P_\textnormal{\tiny Xe}$ (high) via transducer & RS-232 & 20\,s \\
Flow controller display & Teledyne THCD-100~\cite{ref:Teledyne} & 1: Xe purification flow rate & RS-232 & 10\,s \\
Vacuum gauge & Pfeiffer TPR 280~\cite{ref:Pfeiffer} & 1: Pressure insulation vacuum &  RS-232 & 10\,s \\
Universal transducer & Smartec UTI~\cite{ref:uti} & 1: LXe level (long range) & USB & 10\,s \\
Universal transducer & Smartec UTI~\cite{ref:uti} & 3: LXe level (short range) & USB & 10\,s \\
(Safety plugin) & -- & Checks network connection  & -- &  10\,s \\
(Safety plugin) & -- & Checks storage space       & -- & 5\,d \\
\hline
\end{tabular}
\end{table}

The cryostat can currently accommodate up to 15\,kg of liquid xenon, however, the facility can be easily extended to even larger masses. The instruments monitored by the slow control system are listed in Table~\ref{tab:labsetup}. The safety plugin which constantly checks the network connection is installed on the GeMSE server which is using an independent network access point (see Section~\ref{subsec:GeMSE}).

\begin{figure}[htb!]
\centering
\includegraphics[width=\columnwidth]{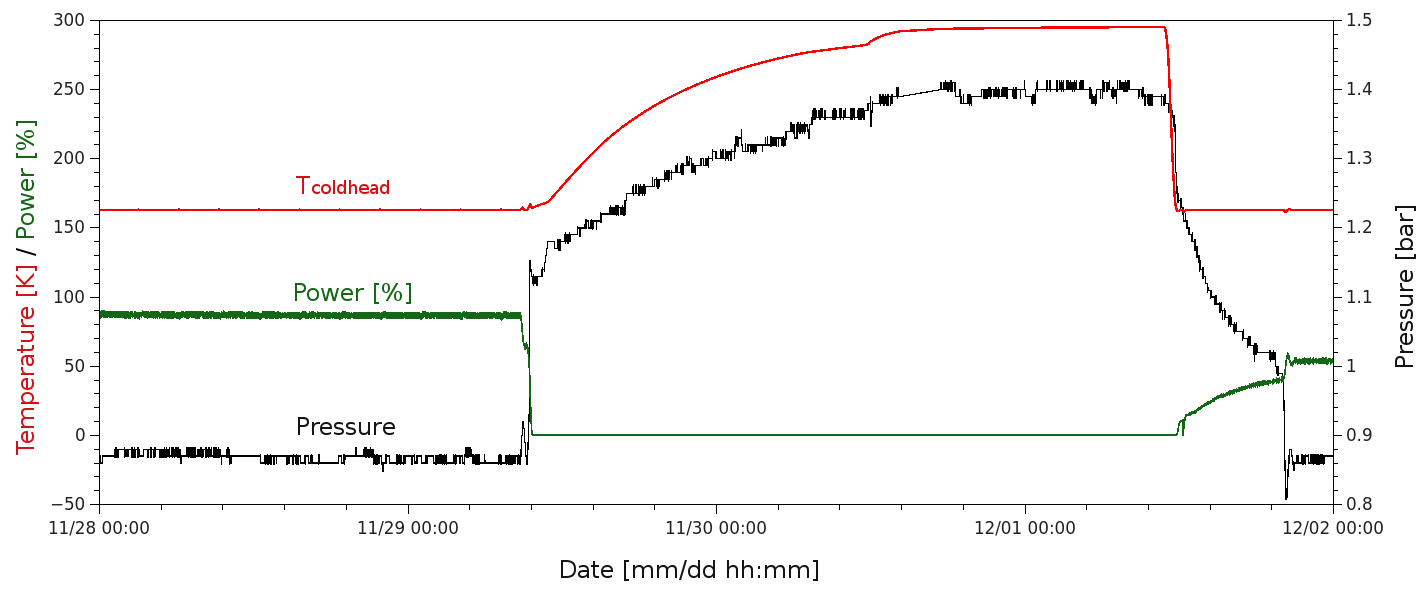}  
\caption{Cold head temperature, heater output power (fraction of 50\,W) and xenon gas pressure recorded over a period of 4\,days. (The liquid xenon temperature is higher than $T_\textnormal{\tiny coldhead}$). The configuration was kept stable for the first 36\,hours, after which the liquid nitrogen cooling and the temperature controlling were  turned off. During the last day shown here, the detector was cooled down again. The data were exported from the Doberman database via the online monitor.}
\label{fig:cryoCon}
\end{figure}

The cryogenic PID controller Cryo-Con~22c from Cryogenic Control Systems Inc.~\cite{ref:cryocon} is the central unit of the system and hence explained in more detail: it measures the temperature of the liquid xenon at two different locations in the cryostat by means of two PT100 sensors. One of them is installed in the liquid xenon and one at the copper cold head on top of the cryostat, which is cooled by liquid nitrogen. To operate the setup with liquid xenon, this temperature is set to the design value (around $-$94$^\circ$C) on the controller, which then powers a heater installed around the cold head to raise its temperature to the set point via a feedback-loop. The set temperatures, actual temperatures and the heater power are measured and transmitted to the slow control system in a single readout process. The setpoints for the feedback-loop can be set and changed via Doberman, while the actual control is done by the controller itself. As an example, Figure~\ref{fig:cryoCon} shows the temperature of the cold head, the xenon gas pressure and the heater power output (fraction of the 50\,W total power available) over a period of 4\,days.

Running the slow control with the instruments and readout periods given in Table~\ref{tab:labsetup}, except the network plugin and UTI transducers, requires a storage space of 4.3\,MB/day. The data consists of raw data, additional Doberman data (e.g., the alarm history) and database overhead (e.g., data type alignment, null bitmap, indices, etc., see~\cite{ref:PostgreSQL}). The Doberman project was developed around this platform and is being successfully used for several months. Its use is indispensable during all cryogenics operations and when the platform is filled with cryogenic liquid xenon, and the availability of its automated alarm system is safety relevant.

\section{Conclusion}\label{sec:Conclusion}

The Doberman slow control system provides a versatile, efficient and reliable way to remotely and automatically control and monitor many experimental parameters at once. It can be easily adapted to very different experimental applications and is hence perfectly suited for small-scale setups at University laboratories, which would otherwise often be operated without a slow control system. Its flexible plugin structure, which provide their data to the main program independently via a data queue allows adding new instruments quickly. Once a plugin is written, it can be reused for identical instrumentation or, with little adjustment, to similar devices. The plugin functionality can also be based on already existing code to communicate to devices or make use of proprietary programs or libraries provided by an instrument manufacturer.

The system is currently successfully deployed on two different experimental projects with very different requirements: a low-background screening facility at a remote location and a detector R\&D platform using cryogenic liquid xenon, see Section~\ref{sec:Examples}. In both cases, about 20\,parameters delivered by 4~or 8~instruments are monitored every 60\,s or 10\,s, respectively. The Doberman system is able to process more than 2500~new parameters from 50~instruments per second. While this currently limits the system's applicability to large projects, it is more than sufficient for most small and medium scale applications.

Doberman is an open source project, which can be downloaded at~\cite{ref::download}.

\section*{Acknowledgments}

This work was supported by the Swiss National Science Foundation (SNF) and the Albert Einstein Center of Fundamental Physics (AEC) at the University of Bern.


\end{document}